\documentstyle[aps,prl,manuscript]{revtex}
\begin{document}
\draft
\newcommand{\beqa}{\begin{eqnarray}}
\newcommand{\eeqa}{\end{eqnarray}}
\newcommand{\beq}{\begin{equation}}
\newcommand{\eeq}{\end{equation}}
\newcommand{\dg}{\dagger}
\newcommand{\sig}{\sigma}
\newcommand{\vektor}[1]{\mbox{\boldmath $#1$}}
\title{One-dimensional Kondo lattice at partial band filling}
\author{Graeme Honner and Mikl\'{o}s Gul\'{a}csi}
\address{Department of Theoretical Physics,
Institute of Advanced Studies \\
The Australian National University,
Canberra, ACT 0200, Australia}
\date{\today}
\maketitle
\begin{abstract}
An effective Hamiltonian for the localized spins in the
one-dimensional Kondo lattice model is derived via a unitary
transformation involving a bosonization of {\it delocalized}
conduction electrons. The effective Hamiltonian is shown to
reproduce all the features of the model as identified in
various numerical simulations, and provides much new
information on the ferro-to-paramagnetic phase transition
and the paramagnetic phase.
\end{abstract}

The Kondo lattice model (KLM) describes the interaction between
a conduction band and a half-filled narrow $f$-band, and is 
thought to capture the essential physics of some of the 
rare earth compounds \cite{Lee}. Although intensively studied 
for two decades, the KLM is still far from being completely 
understood. Even in the simple one-dimensional (1D) model,
and with the conduction band less than half-filled,
there are only two limits in which the behavior has been 
analyzed successfully; in the limit of vanishing conduction 
electron (CE) density, and for antiferromagnetic Kondo couplings
$J>0$, the $f$-electrons ($f$-spins) form a ferromagnetic (FM) 
ground-state \cite{one-electron}; in the strong-coupling limit 
$J \rightarrow \infty$, and for any filling of the conduction 
band, the unpaired $f$-spins are again found to be FM 
\cite{strong-coupling}. The intermediate- to weak-coupling 
regime, away from half-filling but at finite CE density, has 
proved particularly difficult to analyze \cite{strong-coupling}.

From the known limiting behavior 
\cite{one-electron,strong-coupling}, 
together with a consensus of recent numerical simulations
using the density-matrix renormalization-group,
exact numerical diagonalization,
and quantum Monte Carlo \cite{Moukouri,4,Troyer},
a successful theory of the less than 
half-filled 1D KLM will account for the following
ground-state behavior of the $f$-spins: (i) At strong-
to intermediate-coupling the unpaired $f$-spins are
FM at all fillings and show behavior in accord
with the strong-coupling expansion \cite{strong-coupling}.
(ii) As the coupling is lowered, and for finite
CE density, the system undergoes a transition to
a paramagnetic (PM) state, with a filling dependent critical
coupling in the weak to intermediate range.
(iii) At weak-coupling, the system
is characterized by a strong peak in
the $f$-spin structure factor at $2k_{F}$ of the CEs.

In this Letter we derive an effective Hamiltonian
$H_{\mbox{\scriptsize{eff}}}$
from the 1D KLM which reproduces {\it all} the observed
behavior in the intermediate- to
weak-coupling regime. $H_{\mbox{\scriptsize{eff}}}$
treats the $f$-spins exactly while the CEs are treated
using bosonization techniques. The essential new ingredient
in our work is an emphasis on describing delocalized CEs,
as these are responsible for the
observed magnetic behavior of the $f$-spins.
The problem of accessing the intermediate- to
weak-coupling regime nonperturbatively is solved
using a unitary transformation.
The effective Hamiltonian maps to the quantum
random transverse-field Ising spin chain near the
FM-PM boundary, and using extensive work on this
interesting model by Fisher \cite{Fisher}, we can obtain
a vast amount of information on the transition and
the properties of the model near it, as well as
information on the PM phase. 

The Hamiltonian of the 1D KLM is given by
\beq
H=-t \sum_{j\sig} (c_{j\sig}^{\dg}c_{j+1\sig}^{} +
{\rm{H.c.}})+J\sum_{j} {\bf S}_{f j} {\bf \cdot} {\bf S}_{cj}
\label{one}
\eeq
where $t>0$ is the CE hopping,
${\bf S}_{fj} = \frac{1}{2} \sum_{\sig,\sig '}
c_{fj\sig}^{\dg} {\vektor{\sig}}_{\sig,\sig '} c_{fj\sig '}^{}$,
${\bf S}_{cj}=\frac{1}{2}\sum_{\sig,\sig '}
c_{j\sig}^{\dg} {\vektor{\sig}}_{\sig,\sig '} c_{j\sig '}^{}$
and ${\vektor{\sig}}$ are the Pauli spin matrices.  Fermi 
operators $c^{}_{j\sig},c^{\dg}_{j\sig}$ with subscript $f$ 
refer to localized $f$-spins, those without refer to the CEs.
We consider antiferromagnetic Kondo couplings $J>0$ and assume 
the conduction band filling $n = N_{c}/2N < 1/2$ with $N_{c}$ 
the number of CEs and $N$ the number of sites.

From the strong-coupling expansion \cite{strong-coupling},
it is clear that the infinite $J$ on-site spin-singlets,
in which a CE is strictly localized with an $f$-spin, are 
magnetically inert: the strong-coupling FM only appears at large 
but finite $J$ via CE hopping to neighboring unpaired sites, 
with a preferred spin orientation due to broken spin-singlet 
symmetry. The interaction identified in the strong-coupling 
expansion is the Zener double-exchange mechanism. 
This motivates us to introduce a {\it delocalization 
length} $\alpha > a$ ($a$ the lattice spacing) which limits the 
minimum spatial spread of the CEs. The delocalization length 
models the qualitative difference between large $J$ and infinite 
$J$ behaviors, and has its physical basis in the energy gain for 
CE hopping to unpaired $f$-spins whenever $t > 0$. It relates 
to the average spatial spread of the CEs engaged in the 
double-exchange process. For example, the delocalization 
length in the one CE KLM corresponds to the effective spread of 
the spin polaron \cite{one-electron}. For 
simplicity, $\alpha$ will be taken as an average applying 
uniformly to the CEs. It is important to 
emphasize that $\alpha$ limits only the {\it minimum} spread of 
the CEs and does not significantly affect the weak-coupling 
behavior, although it is essential in order to describe the 
strong-coupling FM. 

It is well-known that 1D electrons
may be represented using bosonization techniques.
The Bose description is usually based on the Luttinger
model due to its formal rigor, but this is not
essential. In the present case it is essential {\it not}
to use the Luttinger model, as will become clear.
Two facts, peculiar to 1D, form the basis of
bosonization for realistic 1D systems. The first is Tomonaga's
observation \cite{Tomonaga} that the number fluctuation
operators satisfy Bose-like commutation relations
$ [ \rho_{r\sig}(k),\rho_{r'\sig '}(k') ] =
\delta_{r,r'}\delta_{k,-k'}\delta_{\sig,\sig '}\, rkL/2\pi$
on a weak-coupling long-wavelength subspace, where the
right-moving ($r=+$) and left-moving ($r=-$) number fluctuations
\beqa
\rho_{r\sig}(k) = \sum_{0<rp<\pi/a}
c_{p-\frac{k}{2}\sig}^{\dg}
c_{p+\frac{k}{2}\sig}^{}
\nonumber
\eeqa
with $L=Na$. The second is the fact that these number 
fluctuations generate the 1D state space \cite{Schick}.
The main result from bosonization needed here is the
representation of the Fermi site operators $c_{j\sig}$
in terms of the bosonic number fluctuations $\rho_{r\sig}(k)$.
It is convenient to decompose the site operators into right-
and left-moving components $c_{j\sig}=\sum_{r}c_{rj\sig}$:
\beqa
c_{rj\sig} = \frac{1}{\sqrt N}
\sum_{k_{F}-\frac{1}{2\alpha}<rk<k_{F}+\frac{1}{2\alpha}}
e^{ikja}\,c_{k\sig}
\nonumber
\eeqa
with $k_{F}=\pi n/a$,
and where the momentum cutoff comes from Fourier analysis.
In the Luttinger model the Bose representation
may be formulated as an operator identity \cite{Haldane}.
For the realistic system we must be
satisfied with an approximate representation, but one which
generates asymptotically exact results \cite{Voit}.
(The {\it existence} of the representation is guaranteed by
the completeness of the Bose states.)
In the thermodynamic limit,
\beqa
c_{rj\sig} \approx
{\cal N}(\alpha)e^{irk_{F}ja}\,
e^{i \{ \theta_{\rho}(j)+r\phi_{\rho}(j)
+\sig [ \theta_{\sig}(j)+r\phi_{\sig}(j)] \} / 2}
\label{two}
\eeqa
where the Bose fields for $\nu=\rho,\sig$ are defined by
$\psi_{\nu}(j)  = i (\pi/L)
\sum_{k \ne 0} e^{ikja}
[ \nu_{+}(k) \pm \nu_{-}(k) ] \Lambda(k)/k$,
with $+$ corresponding to the number fields
$\psi_{\nu}=\phi_{\nu}$ and
$-$ to the current fields $\psi_{\nu}=\theta_{\nu}$.
The charge and spin number fluctuations
$\rho_{r}(k)= \sum_{\sig}\rho_{r\sig}(k)$, and
$\sig_{r}(k)=\sum_{\sig}\sig\rho_{r\sig}(k)$.
Eq.\ (2) has the same form as in the Luttinger
model but with one crucial difference.
The even cutoff function $\Lambda(k)$,
satisfying $\Lambda(k) \approx 1$ for $|k| <1/\alpha$ and
$\Lambda(k) \approx 0$ otherwise, is needed in the Bose fields
to ensure that delocalized CEs are
described. The normalization factor ${\cal N}(\alpha)$
depends on both the cutoff and the
cutoff function, and can only be determined asymptotically.
Eq.\ (\ref{two}) will of course fail if it
is used to calculate number operators
$n_{rj\sig}=c_{rj\sig}^{\dg}c_{rj\sig}^{}$. In this case a
Fourier expansion gives
\beqa
n_{rj\sig}=- \frac{a}{4 \pi}
\partial_{x} \{ \phi_{\rho}(j)+r\theta_{\rho}(j)
+\sig [ \phi_{\sig}(j)+r\theta_{\sig}(j) ] \}
\label{three}
\eeqa
to an additive constant.
The separate form for the number operators is manifest
also in the Luttinger model and is accounted for there with
a carefully constructed normal ordering convention and a
prescription for the correct taking of limits \cite{Haldane}.

To derive an effective interaction between the $f$-spins from
the bosonized Hamiltonian (obtained by substituting
Eqs.\ (\ref{two}) and (\ref{three}) in Eq.\ (\ref{one})),
it is sufficient to change to
a basis of states in which the CEs are
coupled to the $f$-spins. This is achieved using
a unitary transformation with
$U=i (aJ/2\pi v_{F})
\sum_{j}S_{fj}^{z}\,\theta_{\sig}(j)$, and where
$v_{F}=2at\sin(\pi n)$.
A variant of this transformation was first used by Emery
and Kivelson for the single-impurity Kondo problem,
and later generalized to the 1D KLM \cite{Emery}.
The usage here is different;
indeed the FM $J^{2}$ term (see Eq.\ (\ref{four}) below),
which $U$ was designed to generate,
is entirely absent in the previous work.
The reason is that a Luttinger model bosonization will miss
any $f$-spin effective interaction which is due to
the {\it non-local} character of the CEs.
Formally, in the Luttinger model the Bose
fields $\phi_{\nu}(j)$ and
$\Pi_{\nu}(j) = -\partial_{x}\theta_{\nu}(j)$
are canonically conjugate and their commutator
strictly vanishes unless they are at the same site.
In our system the fields are smeared over a
range $\alpha$ and their commutator is finite
over roughly $2\pi \alpha$:
$[\phi_{\nu}(j), \Pi_{\nu'}(0)]=
2i\delta_{\nu,\nu'}{\cal J}_{j}(\alpha)$ where
${\cal J}_{j}(\alpha) = \int_{0}^{\infty}
\cos(kja)\Lambda^{2}(k)dk$.
As examples,  a Gaussian
$\Lambda(k)=\exp(-\alpha^{2}k^{2}/2)$ gives
${\cal J}_{j}(\alpha)=(\sqrt{\pi}/2\alpha)
\exp[-(ja/2\alpha)^{2}]$, and the
Luttinger cutoff $\exp(-\alpha |k|/2)$ gives
${\cal J}_{j}(\alpha)
=\alpha/[\alpha^{2}+(ja)^{2}]$.
The Luttinger model
$\delta$-function is obtained by taking $\alpha
\rightarrow 0$ in the last.
The effect of this difference on the transformed Hamiltonian
$\tilde{H} = e^{-U}He^{U}$ is dramatic.
Keeping all terms,
\beqa
\tilde{H} & = &  \frac{av_{F}}{4\pi}\sum_{j,\nu}
\{ \Pi_{\nu}^{2}(j)
+ [\partial_{x}\phi_{\nu}(j)]^{2} \}
-  \frac{a^{2}J^{2}}{4\pi^{2}v_{F}}
\sum_{j,j'}{\cal J}_{j-j'}(\alpha)S_{fj}^{z}S_{fj'}^{z}
\nonumber \\
& + & J{\cal N}^{2}(\alpha)\sum_{j}
[e^{i (1-\frac{aJ}{2\pi v_{F}})
\theta_{\sig}(j)}S_{fj}^{+} + {\rm{H.c.}}]
\{ \cos [K(j)-\phi_{\sig}(j)]
+ \cos [2k_{F}ja+\phi_{\rho}(j)] \}
\nonumber \\
& + & 2J{\cal N}^{2}(\alpha)  \sum_{j}
\sin [K(j)-\phi_{\sig}(j)]
\sin [2k_{F}ja + \phi_{\rho}(j)] S_{fj}^{z}
\label{four}
\eeqa
where $K(j)=-i (aJ/2\pi v_{F}) \sum_{j'}
[\phi_{\nu}(j),\theta_{\nu}(j')]S_{fj'}^{z}$.
A condition for the derivation of
Eq.\ (\ref{four}) is that the cutoff be
not too soft.

The new term in Eq.\ (\ref{four})
is the second. Since $S_{fj}^{z}$ is not
transformed under $U$, it is immediate
that the system is FM at intermediate-coupling at all
fillings. The physical basis for the interaction is quite simple.
A CE spread over more than one lattice site will carry
the same spin over these sites. Due to the term
$J\sum_{j}(n_{rj\uparrow}-n_{rj\downarrow})S_{fj}^{z}$ in
Eq.\ (\ref{one}), this will tend to align the relevant $f$-spins.
This interpretation also makes it clear that the interaction
${\cal J}_{j}(\alpha)$ is short-range provided
$\alpha$ is finite.
We may therefore approximate the FM term
by its nearest-neighbor form $-J_{\mbox{\scriptsize{eff}}}
\sum_{j}S_{fj}^{z}S_{fj+1}^{z}$, $J_{\mbox{\scriptsize{eff}}}=
(a^{2} J^{2}/2\pi^{2} v_{F}) {\cal J}_{1}(\alpha)$.
Although formally this term will give FM at strong-coupling
as well, it is important to recall that the bosonization
describes delocalized CEs. If $J$ is too large then
there will be significant CE localization
and our approximation is less satisfactory. Note that it
is in principle possible to include these effects as well
with a less crude measure of CE delocalization and
with the sum over $j$ in the FM term restricted to sites
containing unpaired $f$-spins only. Such alterations
will not affect our conclusions, except to further
support them.

An effective Hamiltonian for the $f$-spins is obtained
from Eq.\ (\ref{four}) by replacing the CE Bose fields by their
expectation values in the noninteracting ground-state.
This step may be justified for the Bose charge-number field
$\phi_{\rho}(j)$ by noting that at weak-coupling, which is
the only regime where any of the fields affect
Eq.\ (\ref{four}), the charge
structure factor is free electron like \cite{Troyer}. For the
spin fields there is less justification, though note that at
weak-coupling these fields will be relatively smooth and will
enter Eq.\ (\ref{four}) as simple parameters.
Thus while this approximation
may affect the quantitative predictions of the theory, it would
not be expected to affect the qualitative behavior.
(Further evidence for this view was recently provided in
a numerical simulation in which
the same general behavior for the $f$-spins was
seen with $t-J$ interacting CEs \cite{t-J}.)
The effective Hamiltonian is then
\beqa
- H_{\mbox{\scriptsize{eff}}} & = &
J_{\mbox{\scriptsize{eff}}}\sum_{j}S_{fj}^{z}S_{fj+1}^{z}
\nonumber \\
&+& 2J{\cal N}^{2}(\alpha)  \sum_{j}
[\cos K(j)+\cos(2k_{F}ja)]S_{fj}^{x}
\nonumber \\
&+& 2J{\cal N}^{2}(\alpha)\sum_{j}
\sin K(j)\sin(2k_{F}ja)S_{fj}^{z}
\label{five}
\eeqa
and the spin directions have been reversed for later convenience.
Eq.\ (\ref{five}) is our main result.
The remainder of this Letter is concerned with a
brief analysis of $H_{\mbox{\scriptsize{eff}}}$ to show that
it gives all the required behavior. Details will be presented in
a paper to follow \cite{me}.

To describe the destruction of the FM phase, the
$S_{fj'}^{z}$ in $K(j)$ may be replaced by their eigenvalues.
$K(j)$ is then a long-range object which counts the total
$S_{f}^{z}$ to the left of $j$ and subtracts from that the total
$S_{f}^{z}$ to the right. (The effects of the non-Luttinger
bosonization are not important here; $[\phi_{\nu}(j'),
\theta_{\nu}(0)]
\rightarrow i\pi \mbox{sgn}(j')$ at large $j'$.)
Near the FM phase boundary, and in the thermodynamic
limit, it follows that $K(j) \approx 0$ and any transition
is described by the first two terms in $H_{\mbox{\scriptsize{eff}}}$,
with the second term responsible for spin-flips. For incommensurate
fillings, $\cos(2k_{F}ja)$ oscillates unsystematically with respect
to the lattice. The large values $\cos(2k_{F}ja) \approx 1$ which
are responsible for spin flips, are then widely separated. Following
analogous treatments in spin-glasses \cite{Abrikosov}, this behavior
is well-described by taking
$\cos(2k_{F}ja)$ as a random variable.
The factor multiplying $S_{fj}^{x}$
in Eq.\ (\ref{five}) is then replaced by $h_{j}$,
where $h_{j}$ is drawn independently from the
displaced $\cos$ distribution $\rho(h)dh$ where
$\rho(h) = (1/C\pi) \{ 1-[(h/C)-1]^{2} \}^{-1/2}$
and $C=2J{\cal N}^{2}(\alpha)$.
Note that fluctuations in the Bose charge-number fields
$\phi_{\rho}(j)$ offer further support for this
interpretation. The behavior of the $f$-spins
at and near the destruction of the FM phase
is then governed by the quantum
random transverse-field Ising spin
Hamiltonian $H_{\mbox{\scriptsize{crit}}}
=-J_{\mbox{\scriptsize{eff}}}\sum_{j}
S_{fj}^{z}S_{fj+1}^{z}-\sum_{j}h_{j}S_{fj}^{x}$. Using
extensive real space renormalization-group work on this model
by Fisher \cite{Fisher} (to whom we refer the reader for
details), we determine the location of the
quantum critical line describing the order-disorder
transition at
\beq
\frac{J}{t}=
\frac{4\pi^{2}{\cal N}^{2}(\alpha)}{a {\cal J}_{1}(\alpha)}
\sin(\pi n)
\label{six} \; .
\eeq
The numerical predictive powers of
$H_{\mbox{\scriptsize{eff}}}$ are
restricted by lack of knowledge of ${\cal N}(\alpha)$.
We would like to emphasize that such problems beset
{\it any} bosonization description  in which physical
quantities are found to depend on this factor,
and are not due to our particular bosonization.
Accordingly, the coefficient of $\sin(\pi n)$ in Eq.\ (\ref{six})
is used as a fitting parameter to numerically obtained
critical points \cite{Moukouri,4,Troyer}. A good fit
is obtained with $J/t = 2.5 \sin(\pi n)$ as shown in
Fig.\ 1. Note that this ignores any functional dependence of
$\alpha$ on $J$ or $n$. 

For the following discussion, it is convenient to introduce
a measure of deviation from criticality $\delta \propto
\ln [ 2\pi^{2} {\cal N}^{2}(\alpha) v_{F} /
a^{2}J{\cal J}_{1}(\alpha)] $ \cite{Fisher}, which for the
obtained fit is $\delta \propto \ln [2.5 t \sin(\pi n) / J ]$.

The behavior described by $H_{\mbox{\scriptsize{crit}}}$
is simply understood in terms of {\it clusters} of ordered
$f$-spins. Reducing $J$ from intermediate values in the FM
phase, the infinite cluster characterizing strong FM is
broken up into several large clusters as the quantum
fluctuations $h_{j}$, controlled by the spin-flip interactions,
become stronger. The individual clusters are the spin polarons. 
The system is weakly ordered, and exists
for $-0.7<\delta < 0$ with the boundary determined by
$J_{\mbox{\scriptsize{eff}}}=\mbox{max}\{ h_{j} \} $,
as shown in Fig.\ 1.
This is not a true transition line, but rather marks
the onset of a Griffiths phase \cite{Griffiths} characterized
by singularities in the free energy over the whole range of $\delta$.
For small $\delta$ the correlation length is
$\xi \sim \delta^{-2}$,
beyond which the system is ordered. The spontaneous
magnetization $M_{0} \propto | \delta |^{\beta}$ with $\beta=
(3 - {\sqrt{5}})/2 \approx 0.38$, while for small applied
fields $H$ the magnetization $M(H) \propto M_{0}[1+
{\cal{O}}(H^{2|\delta|}\delta \ln H)]$; the susceptibility
is infinite with a continuously variable exponent.
The mean correlation function $\overline{\langle S_{fj}^{z}
S_{fj+x}^{z}\rangle} - M_{0}^{2} \propto
|\delta|^{2\beta} (\xi/x)^{5/6} e^{-x/\xi}
\exp[-3(\pi x/\xi)^{1/3}]$ for $x \gg \xi$ and where the
averaging is over $\rho(h)$ \cite{Fisher}.

Further lowering $J$, we reach the true phase transition
Eq.\ (\ref{six}). The correlation length is infinite, the
magnetization $M(H) \propto |\ln H|^{-\beta}$ for small $H$, and
the mean correlation function $\overline{\langle S_{fj}^{z}
S_{fj+x}^{z}\rangle} \propto x^{-\beta}$.

Immediately below the critical line ($\delta >0$), the system
presents a weakly disordered Griffiths phase. The remaining
clusters occupy a small fraction of the system length but
``think'' that they are still in the ordered phase; their
magnetization $\delta^{\beta}$ per unit length is identical
to $M_{0}$ of the weakly ordered phase. These remaining rare
clusters dominate the low-energy physics. The magnetization
$M(H) \propto \delta^{\beta} \{ H^{2\delta}
[\delta \ln(1/H) + {\rm{const.}}] + {\cal{O}}
[H^{4\delta} \delta \ln(1/H)] \}$; thus
$M(H)$ has a power law singularity with a continuously
variable exponent $2\delta$; as in the weakly ordered
phase the susceptibility is {\it infinite}.
The mean correlation function decays less rapidly than in
the ordered phase, but takes the same form
$\overline{\langle S_{fj}^{z} S_{fj+x}^{z} \rangle}
\propto \delta^{2\beta} (\xi /x)^{5/6} e^{-x/ \xi}
\exp[-3/2 (\pi x/\xi)^{1/3}]$ for $x \gg \xi = 1/\delta^{2}$.
According to
$H_{\mbox{\scriptsize{crit}}}$, the weakly disordered
Griffiths phase extends down to $J=0$. However, as the
disorder increases, the third term in
$H_{\mbox{\scriptsize{eff}}}$ is no longer negligible.
At very low $J$, the last two terms in
$H_{\mbox{\scriptsize{eff}}}$ will dominate; this
corresponds to free spins in a field with dominant
correlations at $2k_{F}$ of the conduction band,
and is responsible for the observed peak in the $f$-spin
structure factor \cite{Moukouri,4,Troyer}. No clusters remain.
This strongly disordered conventional PM phase is
indicated schematically in Fig.\ 1.

In summary we have derived an effective Hamiltonian for the
$f$-spins in the 1D KLM which reproduces all the behavior
seen in numerical simulations in the intermediate-
to weak-coupling regime: (i)
$H_{\mbox{\scriptsize{eff}}}$ presents a FM phase at
intermediate-coupling due to ``forward'' scattering by
delocalized CEs, and is consistent with known limiting behavior
\cite{one-electron,strong-coupling}. (ii) As $J$ is
lowered this phase is gradually disordered due to spin-flip
interactions between the CEs and the $f$-spins.
A sharp quantum order-disorder transition occurs to a
PM phase at a critical coupling given in Eq.\ (\ref{six}).
(iii) The backscattering interactions leave a residue correlation
at $2k_{F}$ in the $f$-spins at weak-coupling.

This work was supported by the Australian Research Council.

\figure{Fig.\ 1.\ \
Ground-state phase diagram of the 1D KLM. The solid
(critical) line is from Eq.\ (\ref{six}) with
$4\pi^{2}{\cal N}^{2}(\alpha)/a {\cal J}_{1}(\alpha)$ used as
a fitting parameter to numerically determined points: square
is density-matrix-renormalization-group data of a 75 site chain
from Ref.\ \cite{Moukouri}; diamond is the quantum Monte Carlo
data for a 24 site system from Ref.\ \cite{Troyer}; open
circles are the exact numerical diagonalization data
for the 8 site chain from Ref.\ \cite{4}. The dashed lines
separate conventional strongly ordered (FM)/disordered (PM)
phases from their weak (Griffiths phase) counterparts.}
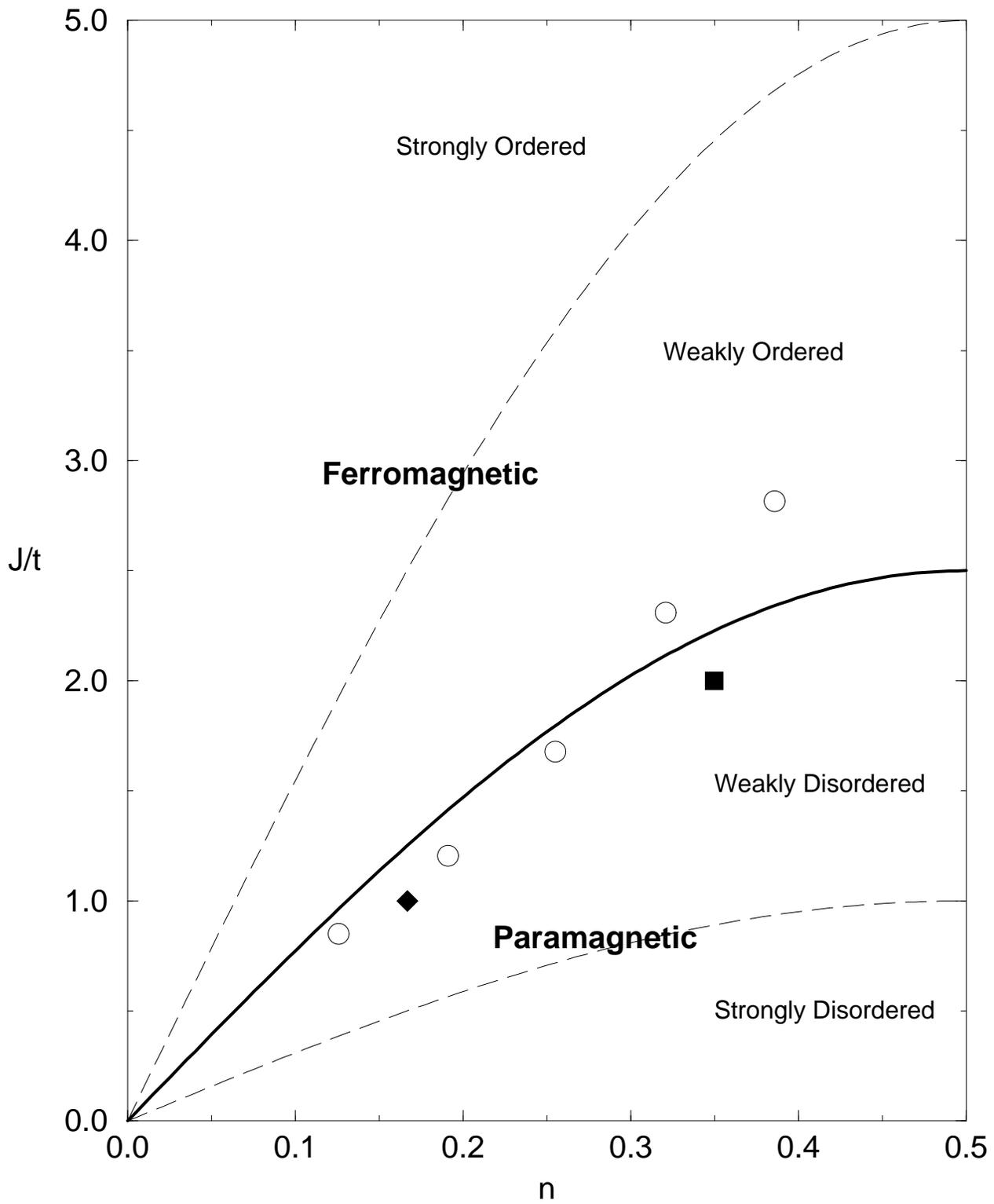
\end{document}